\documentclass{article}
\usepackage{spconf,graphicx}

\title{Multi-Channel Automatic Speech Recognition Using Deep Complex Unet}
\name{Yuxiang Kong$^{1,2}$, Jian Wu$^1$, Quandong Wang$^2$, Peng Gao$^2$, Weiji Zhuang$^2$, Yujun Wang$^2$, Lei Xie$^{1\dag}$}
\address{$^1$Audio, Speech and Language Processing Group (ASLP@NPU), School of Computer Science,\\ Northwestern Polytechnical University, Xi’an, China\\
  $^2$Xiaomi Inc., Beijing, China}
\begin{document}
%
\maketitle
\begin{abstract}
The front-end module in multi-channel automatic speech recognition (ASR) systems mainly use microphone array techniques to produce enhanced signals in noisy conditions with reverberation and echos. Recently, neural network (NN) based front-end has shown promising improvement over the conventional signal processing methods. In this paper, we propose to adopt the architecture of deep complex Unet (DCUnet) - a powerful complex-valued Unet-structured speech enhancement model -  as the front-end of the multi-channel acoustic model, and integrate them in a multi-task learning (MTL) framework along with cascaded framework for comparison. Meanwhile, we investigate the proposed methods with several training strategies to improve the recognition accuracy on the 1000-hours real-world XiaoMi smart speaker data with echos. Experiments show that our proposed DCUnet-MTL method brings about 12.2\% relative character error rate (CER) reduction compared with the traditional approach with array processing plus single-channel acoustic model. It also achieves superior performance than the recently proposed neural beamforming method.
\end{abstract}
\begin{keywords}
Multi-channel speech recognition, robust speech recognition, deep learning, deep complex unet
\end{keywords}
\renewcommand{\thefootnote}{\fnsymbol{footnote}}
\footnotetext{\dag  Lei Xie is the corresponding author.}
\section{Introduction}
\label{sec:intro}

Automatic speech recognition (ASR) has made great progress in model architectures and optimization criteria~\cite{povey2016purely,prabhavalkar2018minimum,graves2006ctc,chan2016las,vaswani2017transformer} in close-talk scenarios. However, in  far-field acoustic environment, the speech recognition accuracy will degrade significantly due to the corrupted speech caused by reverberations, directional noises, and sometimes echos from smart devices. Thus the front-end processing, including speech enhancement and dereverberation techniques, becomes an indispensable module~\cite{arrays2001signal} in distant speech recognition. 

A common far-field ASR system can be roughly divided into two parts, a multi-channel single processing front-end and a single-channel acoustic model in the back-end~\cite{hain2011transcribing}. Traditional front-end incorporates microphone array techniques which make use of spatial information from multi-channel signals and produce the enhanced signal. Then output signal is fed into the back-end module which performs decoding with acoustic and language models. In this way, the optimization objectives of the front-end and the back-end may be inconsistent, where improving speech quality may not necessarily lead to lower word error rate (WER). With the rapid development of neural networks (NN), the multi-channel signal processing strategies incorporating neural techniques have been proposed~\cite{sainath2017google} and joint optimization of the front-end and the acoustic model can be easily implemented. It mitigates the difference between target of enhancement, which contributes to SNR as well as acoustic likelihood, and ASR which is eager for high accuracy~\cite{mcdonough2008distant,seltzer2008bridging}.

Beamforming~\cite{benesty2008microphone} is a typical multi-channel speech enhancement method in conventional signal processing, which constraints signal distortion on source direction and suppresses noises from the other directions~\cite{benesty2007springer}. Previous studies can be roughly categorized into fixed beamforming (FB) and adaptive beamforming. There has been much interest in introducing neural network architecture to perform FB as it can benefit ASR without introducing much latency. Adopting neural beamforming, the geometry dependent filtering coefficients are learned in a data-driven manner~\cite{xiao2016xiaoxiong,sainath2017google} instead of handcrafted design. In~\cite{wu2019amazon}, initialization and beam selection were explored to improve the accuracy of speech recognition. An alternative way to integrate adaptive beamforming with acoustic modeling is based on the mask-based beamforming techniques~\cite{heymann2017beamnet,heymann2018performance,menne2018speaker}. The adaptive filtering weights, e.g., MVDR, are calculated depending on the estimation of the time-frequency mask, and the mask network is jointly optimized with the acoustic model, only using the cross-entropy criteria.  Those approaches often follow a cascade structure, where a speech enhancement network is followed by a conventional acoustic model, and the enhancement module is only adjusted with the acoustic model's criteria. Another way to improve speech recognition accuracy is to take enhancement network as an auxiliary task and recognition as the main task in the form of multi-task learning (MTL)~\cite{pironkov2016multi}. In this way, the network can effectively use a shared representation to learn both classifying and enhancing the acoustic observations at the same time. Various model structures~\cite{giri2015improving,zhang2017attention,chen2015speech,zhao2019multi} have been explored in this direction.

Recently, the neural speech enhancement systems have begun to handle the phase issues through time-domain processing~\cite{pandey2019new,wu2019time}, complex spectrogram prediction or masking~~\cite{tan2019learning}, as well as performing phase modification~\cite{yin2019phasen,hyx2020dccrn}, which boost enhancement quality significantly and achieve extraordinary performance. Among them, deep complex Unet (DCUnet)~\cite{choi2019dcunet} can produce more precise phase estimation using a new polar coordinate-wise complex-valued masking method. It directly deals with complex-valued spectrograms by combining the advantages of both complex networks and U-Net~\cite{ronneberger2015u}, achieving state-of-the-art performance for speech enhancement.

In this paper, in order to improve the performance of multi-channel ASR, we adopt the architecture of the DCUnet to deal with multi-channel signal enhancement in the frequency domain. We look into whether the recent advances of speech enhancement on complex spectrogram can benefit multi-channel ASR. Specifically, we investigate two different front-end processing methods based on DCUnet. One is to compose the DCUnet and the acoustic model in a cascade structure (DCUnet-CAS). And another one is to integrate them in the MTL framework (DCUnet-MTL), where the main task is speech recognition and taking enhancement as an auxiliary task. In details, in our MTL framework, the encoder of the DCUnet is shared by both tasks and the decoder is only used for the enhancement branch. In real-world XiaoMi smart speaker Mandarin dataset which contains 1000 hours of 2-microphone signal as well as a reference channel of echo, the DCUnet-MTL strategy impressively brings 12.2\% relative character error rate (CER) reduction compared to the baseline system -- speech processing using a conventional pipeline and decoding using a single-channel acoustic model.  It also achieves better performance than a neural fixed beamforming approach similar to the one in~\cite{wu2019amazon}.

\section{Proposed methods}
\label{sec:format}

The data we used in this paper is collected by the XiaoMi smart speaker\footnote{https://www.mi.com/aispeaker-touch?cfrom=search} in real scenes. The device is equipped with two microphones and one build-in loudspeaker, collecting signals with three channels: the first two channels are received signals which may be interrupted by acoustic echoes, and the third one is the reference channel (for echoes) recording original signals from the build-in loudspeaker.

\subsection{Multi-channel DCUnet}
\begin{figure}[t]
    \centering
    \includegraphics[width=\linewidth]{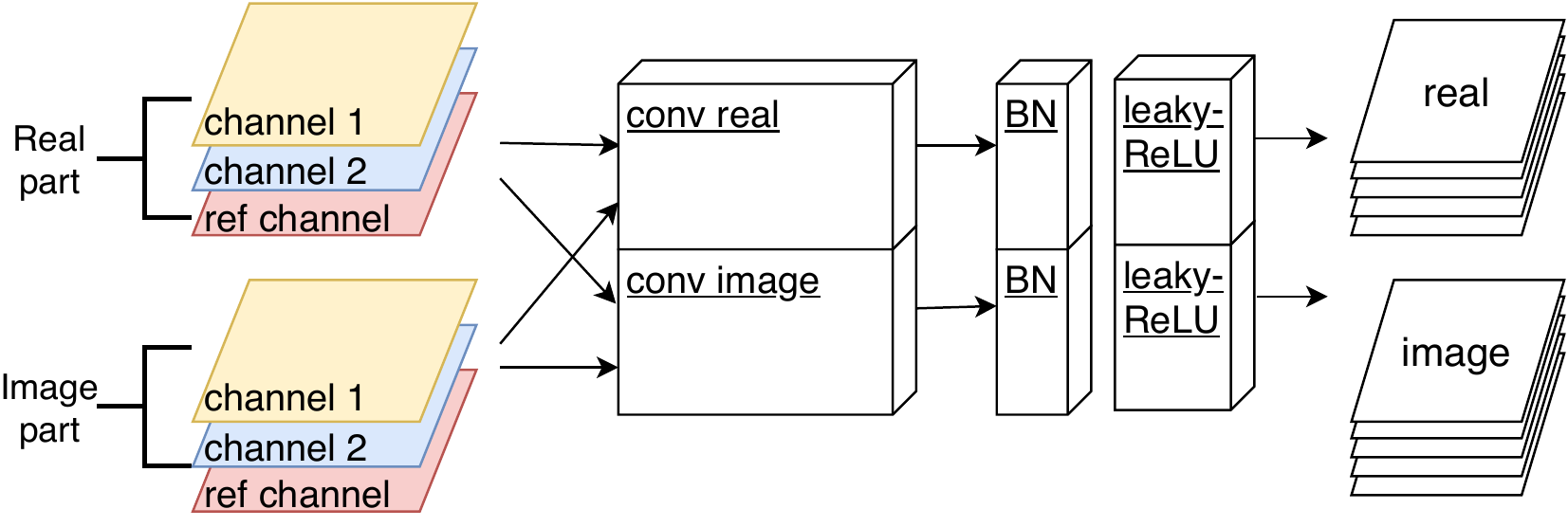}
    \caption{Input layer of multi-channel DCUnet}
    \label{fig:mcdcunet}
\end{figure}

DCUnet~\cite{choi2019dcunet} is an extension of the Unet structure proposed in image segmentation using complex-valued convolutional encoders and decoders. The encoder and decoder are stacked by several blocks which contain 2D (de)convolution layers, batch normalization (BN) and leaky rectified linear units (leaky-ReLU).
The multi-channel version we used in this paper takes the short-time Fourier transform (STFT) of multi-channel signals as the input features. The structure of the input layer is shown in Fig.\ref{fig:mcdcunet}, where the STFT of multi-channel signal is divided into real and imaginary parts and convolves with the corresponding filters. For each channel $c$, the complex-valued convolutional operation can be interpreted as two real-valued ones, following the formula:
\begin{equation}
\mathbf{W} \circledast \mathbf{X}_c = 
\begin{bmatrix}
\mathbf{W}_i \\
\mathbf{W}_r
\end{bmatrix} \circledast
\begin{bmatrix}
\mathbf{R}_c \\
\mathbf{I}_c
\end{bmatrix}
= 
\begin{bmatrix}
\mathbf{W}_i * \mathbf{R}_c - \mathbf{W}_r * \mathbf{I}_c \\
\mathbf{W}_i * \mathbf{I}_c + \mathbf{W}_r * \mathbf{R}_c
\end{bmatrix}
\end{equation}
where $\circledast$ and $*$ denote complex-valued and real-valued convolution, respectively. $\mathbf{W} = [\mathbf{W}_r, \mathbf{W}_i]^T$ are complex-valued convolution filters and  $\mathbf{X}_c = [\mathbf{R}_c, \mathbf{I}_c]^T, c \in [1, 2, \text{ref}]$ are STFT of the signal at channel $c$.
Batch normalization is applied on real and image channels independently, excluding the final layer of the decoders.

Different from the original structure designed for single-channel speech enhancement, the multi-channel DCUnet does not estimate complex ratio mask (cRM)~\cite{williamson2015complex} but predicts the complex spectrogram directly because the training data is collected in real-world environments and lack of ground-truth signals. The audio after the signal processing steps are used as the supervision of the DCUnet and the scale issues may exist. Thus the mask-based loss function can not be optimized stably. We use the mean square error (MSE) between the magnitude of the supervision $\mathbf{M}_{\text{sup}}$ and the prediction output $\mathbf{O}_{\text{DCUnet}}$ as the objective function as shown in Eq.~(2), considering that the phase is always neglected during the feature extraction stage in the recognition task.
\begin{equation}
    \mathcal{L}_\text{enh} = \Vert\mathbf{M}_{\text{sup}} - |\mathbf{O}_{\text{DCUnet}}| \Vert_F^2
\end{equation}

\subsection{Cascade Structure}
\begin{figure}[t]
    \centering
    \includegraphics[width=0.50\linewidth]{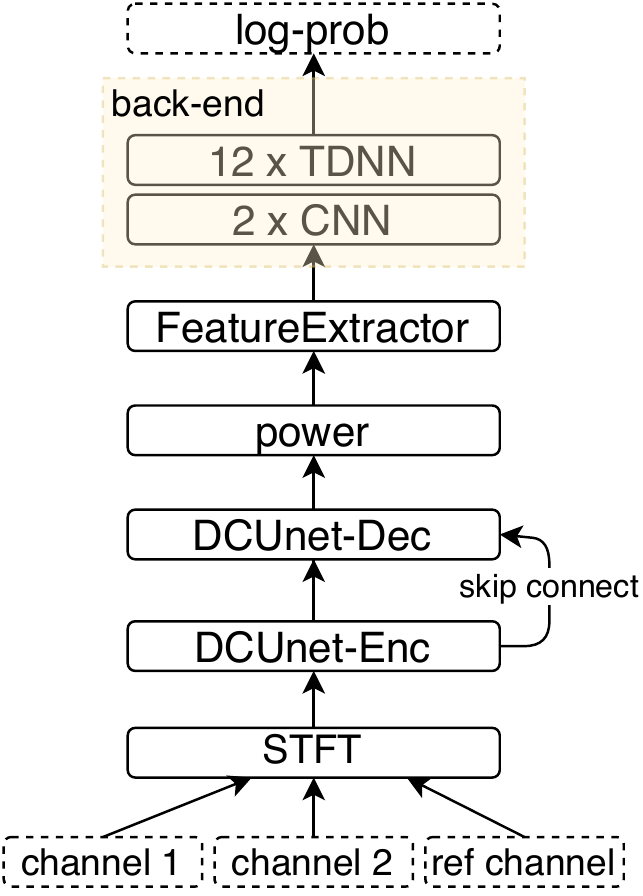}
    \caption{The whole DCUnet structure followed by back-end}
    \label{fig:fulldcunet}
\end{figure}

One straightforward way to joint the multi-channel DCUnet with the acoustic model is to compose them in a cascaded structure, as shown in Fig.~\ref{fig:fulldcunet}. We expect the DCUnet to learn the spectrogram representation that benefits speech recognition without signal supervision and fully depends on the optimization of the back-end. At the beginning, STFT is applied to the input signals (1,2 and reference channels) to obtain complex spectrogram. Then the DCUnet predicts the complex spectrogram of each channel. After that, the power spectrogram is calculated and the feature extraction module computes the log mel-filter bank (log-FBank) features. The acoustic model used in the back-end contains a 2-layer CNNs and a 12-layer time delay neural networks (TDNN). The log-softmax layer is applied in the last to obtain the log-probabilities. 

\subsection{MTL Structure}

\begin{figure}[th]
    \centering
    \includegraphics[width=0.90\linewidth]{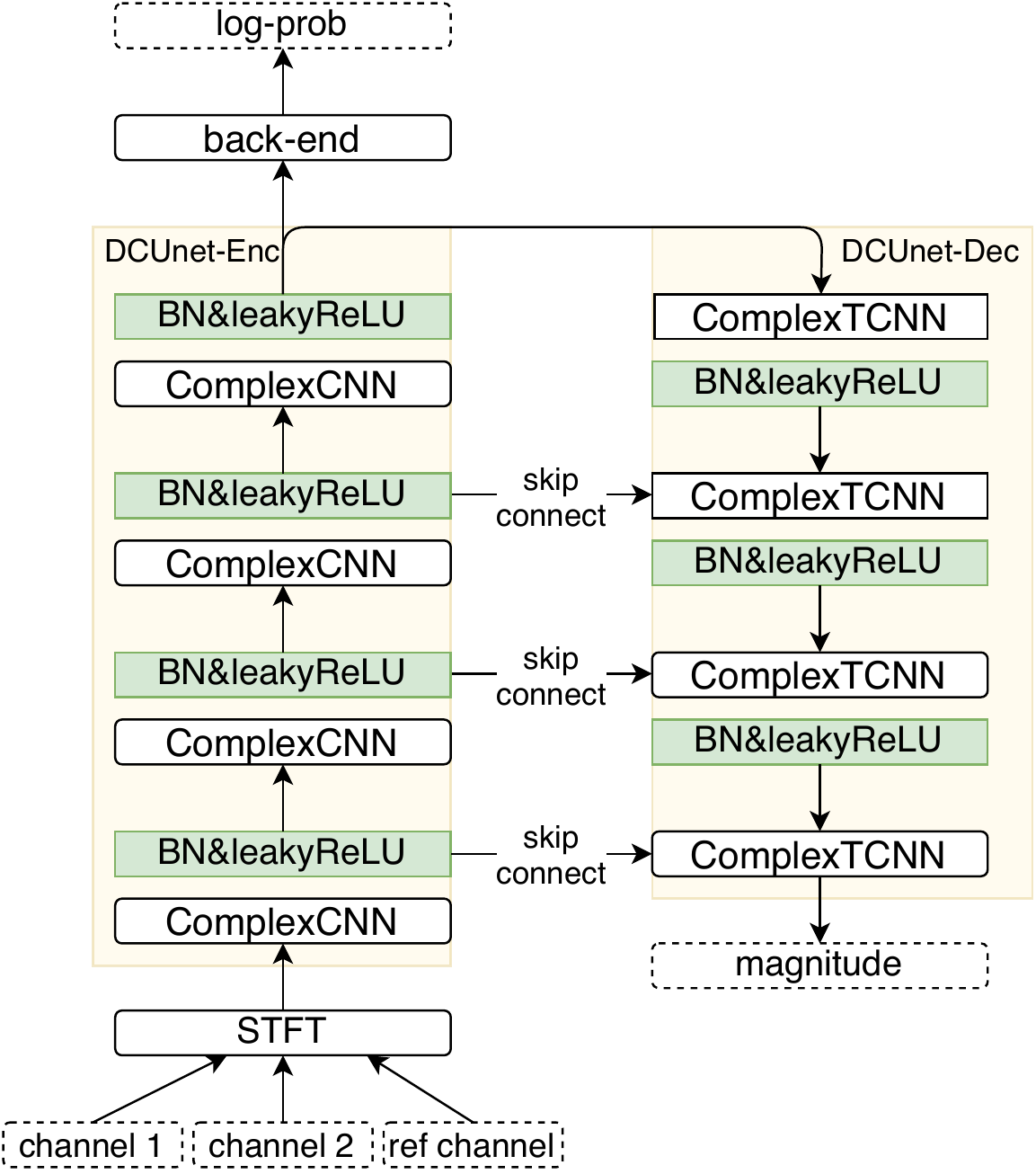}
    \caption{DCUnet-MTL framework}
    \label{fig:dcunet}
\end{figure}

Along with the enhancement-recognition cascade structure, we can also use a multi-task learning (MTL) framework, where the main task is speech recognition and the auxiliary task is speech enhancement. We make the encoders shared by enhancement decoders and recognition network in order to extract the representation of the input signals that can benefit both tasks. The framework is shown in Fig.\ref{fig:dcunet}, where TCNN is the abbreviation of transposed convolution layers.

 It is worth noting that the features generated by DCUnet encoder are still in complex values, while the first layer of the back-end is a CNN of real-value. In order to join the complex CNN and the real-valued one, real and imaginary parts of the complex CNN outputs are regarded as two different channels in order to be compatible with the real-valued CNNs. For example, the output of DCUnet encoder consists of 8 complex feature maps, and we send it to the real-valued CNN as 16 real-valued feature maps.

The structure of the DCUnet in  cascaded and MTL framework are same, where the encoder contains 4 complex convolutional layers and the decoders consists of 4 complex deconvolutional layers. Specifically, the number of the output channels for the first three convolutional layers are 16, 32, 64 with the kernel size $7\times 5$,  respectively. And the last layer contains 64 filters with the kernel size $5\times 3$.

\subsection{Model Training}
In DCUnet-MTL structure, the gradients back-propagated from the two outputs are weighted by $\beta$ and $1-\beta$ for enhancement and recognition tasks respectively.  Specifically, the objective function for the DCUnet-MTL network can be written as:

\begin{equation}
\mathcal{L} = \begin{cases}
(1 - \beta) \mathcal{L}_{\text{asr}} + \beta \mathcal{L}_{\text{enh}} &\text{$t \leqslant T_{\text{enh}}$},\\
\mathcal{L}_{\text{asr}} &\text{$t > T_{\text{enh}}$}
\end{cases}
\end{equation}
where $\mathcal{L}_{\text{asr}}$ and $\mathcal{L}_{\text{enh}}$ stands for recognition and enhancement loss functions, respectively. $\beta$ is a weight parameter between 0 and 1. $T_{\text{enh}}$ represents the number of enhancement task iterations and $t$ indicates the current iteration number in the training progress. The enhancement task will only iterates for the first $T_{\text{enh}}$ times, and after that the recognition criterion will take over the training. The end-to-end lattice-free MMI criterion~\cite{hadian2018improving,povey2016purely} is served as the objective function on the recognition task in our paper.

The training of the enhancement branch requires the noisy-clean data pairs. As explained in Section 2.1, we use the data generated by signal pre-processing steps, which consists of the Wiener filter~\cite{benesty2007springer} based automatic echo cancellation (AEC) followed by delay and sum (DS) beamforming\footnote{https://github.com/xanguera/BeamformIt}, as the supervision. Specifically, we performed AEC on the first two channels (channel 1, 2) using the third channel signal as the echo reference. Then a simple DS beamformer is adopted to form the final enhanced signal, which is treated as the ``target signal" for training of the DCUnet. Obviously it's not the ideal supervision, so we only enable $\mathcal{L}_{\text{enh}}$ for several iterations.

In order to further improve the performance, we use a well-trained DCUnet as initialization. Specifically, the full-structured DCUnet for the initialization purpose is firstly trained on the same dataset, and the DCUnet part in the two structures (cascaded and MTL) is initialized with this well-trained DCUnet. We find that such initialization benefits the performance for both structures. Besides, extra gains can be found by inserting a dropout layer before the back-end acoustic model.

\section{Experiments}
\label{sec:pagestyle}

\subsection{Dataset}
All experiments are performed on the real-world Mandarin data from XiaoMi smart speaker. The device is equipped with two receivers in horizontal dimension, and the distance between the two receivers is about 7cm. A build-in loudspeaker is located in the middle of the two receivers. Before speech communications triggered by a hot word, the smart speaker may be playing music or speech responses, or it keeps silence. Once waken up, it will turn the speaker volume down rather than off; at the same time, it will give a speech response, such as ``I'm here" in Chinese. The interactions with the smart speaker are unconstrained. The 16kHz training and testing data amounts to approximately 1000 hours and 35 hours, respectively.

\subsection{Baseline System}
In order to be comparable with the traditional technique solutions, we use the audio data processed by AEC followed by the DS beamforming to train the acoustic model, which is served as the single channel baseline. The 80-dimensional log-FBanks are extracted as the input features, with 25ms window size and 10ms frame shift. 
The acoustic model consists of 2-layer real-valued CNN followed by 12-layer TDNN. Specifically, the first convolutional layer with 64 filters of size $5\times3$ is used and the second one is of $3\times3$ with 128 filters. The first 11 TDNN layers, each has 1024 hidden nodes and followed by a batch normalization layer and the rectified linear units (ReLU). The last layer has 2888 units with log-softmax applied to the output.
\begin{figure}[t]
        \centering   
        \includegraphics[width=0.35\linewidth]{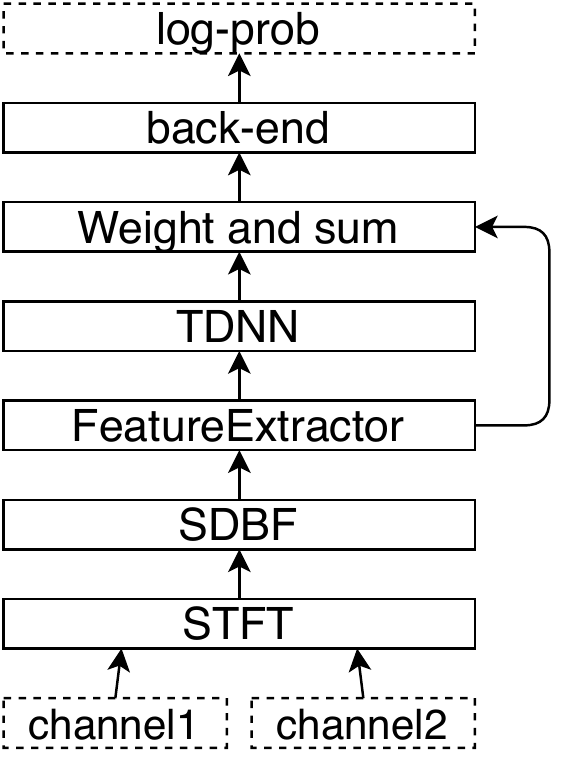}   
        \caption{Neural fixed beamforming for comparison}
        \label{fig:sdbf}   
\end{figure}

Besides, we try a typical neural fixed beamforming (NNFB) method similar to~\cite{sainath2017google,wu2019amazon} as the multi-channel acoustic model baseline. The architecture contains a fixed beamformer layer, which aims to enhance source signals on different look directions. As suggested in~\cite{wu2019amazon}, we initialize the filter coefficients using the SuperDirective (SD) beamformers. For beam selection, we employ a TDNN layer to predict the weight on each direction and the sum of the weighted directional log-FBank features is used as the input of the acoustic model, as shown in Fig.~\ref{fig:sdbf}. In our experiments, we use 8 different directions for the fixed beamformer layers and the two-channel inputs are processed after the AEC steps, considering that the neural beamforming method can not handle echo issues well.

\subsection{Comparison in Different Models}

The comparison of the four systems -- single-channel ASR, NNFB, DCUnet-MTL and DCUnet-CAS, is displayed in Table~\ref{tab:structures}. It is clear that the NNFB provide higher accuracy than single-channel ASR because the trainable beamforming filters can be optimized jointly with the input features and acoustic network, which is more likely to achieve the optimal solution for recognition task. For the DCUnet-MTL structure, we test two values for the iteration number $T_{\text{enh}}$ of the enhancement task. Experiments show that the result of training without the regularization of the enhancement task ($T_{\text{enh}}=0$) is better and we observe that the ASR branch converges faster than the enhancement branch during training progress. In order to achieve better performance, we train the DCUnet front-end separately and get those DCUnet based structures initialized with the well-trained model, which will be discussed in the next section.

\begin{table}[htb]\vspace{-5pt}
  \caption{Comparison of different structures}
  \label{tab:structures}
  \centering
  \begin{tabular}{ccccc}
   \toprule
    \textbf{Method} & \textbf{Prep} & \textbf{CH} & \textbf{$T_{\text{enh}}$}     & \textbf{CER (\%)}  \\
    \midrule
    Baseline    & AEC + FB & 1 & -   & 15.08      \\
    NNFB    & AEC & 2 & -     & 14.67 \\
    DCUnet-MTL   & - & 3 & 0 & 14.67 \\
    DCUnet-MTL   & - & 3 & 6   & 15.42 \\
    DCUnet-CAS  & - & 3 & - & 16.51 \\
    \bottomrule
  \end{tabular}\vspace{-10pt}
\end{table} 

\subsection{Initialization and Dropout}

As is demonstrated in Table~\ref{tab:dropout}, recognition accuracy improves after better initialization strategy is used and the DCUnet-MTL structure can get higher accuracy with the regularization of the enhancement branch. Considering the supervision we used is not ideal for enhancement task, moderate guidance of enhancement can be a better choice. We find that iterating the enhancement and recognition branch for 3 and 6 epochs shows the best CER 14.03\% on this task. DCUnet-CAS model can also benefit from the better initialization, improving the result from 16.51\% to 14.09\%. 

Another conclusion could be drawn from the table is that the extra gains would be got if we insert a dropout layer before acoustic model in our proposed structure, especially after getting DCUnet better initialized. And the best CER is pushed to 13.48\% with the usage of the dropout. However, single channel network and NNFB cannot get gains from that.

\begin{table}[htb]\vspace{-5pt}
  \caption{CER (\%) of initialization \& dropout layer}
  \label{tab:dropout}
  \centering
  
  \begin{tabular}{ccccc}
   \toprule
   \textbf{Method}   & \textbf{Initialization}    & \textbf{$T_{\text{enh}}$}  & \multicolumn{2}{c}{\textbf{Dropout}}  \\
   & & & $\times$ & $\checkmark$ \\ \midrule
   Baseline     & Random    & -  & 15.08   & 16.32 \\ \midrule
   NNFB     & Random    & -  & 14.67   & 14.84 \\ \midrule
    DCUnet-MTL   & Random    & 0 & 14.67 & 14.32 \\
     & DCUnet   & 0  & 14.39 & 13.74 \\
     & DCUnet   & 3  & \textbf{14.03}   & \textbf{13.48} \\
     & DCUnet   & 6  & 14.16   & 13.64 \\ \midrule
    DCUnet-CAS     & DCUnet    & -  & 14.09   & 13.82 \\
    \bottomrule
  \end{tabular}\vspace{-10pt}
  
\end{table}

\subsection{Weight of Multi-task Learning}

We further tune the weights of recognition and enhancement tasks, while fixing the iteration times of the enhancement branch with the usage of the dropout layer and model initialization. The result is shown in Table~\ref{tab:weight}, where we can see that the structure can benefit from the enhancement task with higher weight in the first 3 iterations. Finally the CER reaches the lowest 13.23\% on our test set.

\begin{table}[htb]\vspace{-5pt}
  \caption{Comparison of different weights}
  \label{tab:weight}
  \centering
  \begin{tabular}{cccc}
   \toprule
    \textbf{Method}     &\textbf{$T_{\text{enh}}$} &\textbf{$\beta$} &\textbf{CER (\%)}  \\
    \midrule
    DCUnet-MTL   & 3    & 0.2    & 13.42 \\
       & 3    & 0.5    & 13.39 \\
       & 3    & 0.8    & \textbf{13.23} \\
    \bottomrule
  \end{tabular}\vspace{-15pt}
\end{table} 

\subsection{CER in Detail}
In order to observe the results closely, we further divided the test set into four subsets, one with echos in the reference channel and the others are divided according to the signal-to-noise ratio (SNR). All the tested utterances start with the speech response of the smart speaker, which last around 0.8 second. Roughly speaking, in test set, data with echos takes up 23.5\%. And for the rest non-echo part, data with SNR lower than 5dB occupies 14\%, and data with SNR between 5dB and 15 dB accounts for 32\%. The reset 30.5\% of test set is the data whose SNR is higher than 15dB.


\begin{table}[htb]\vspace{-6pt}
\setlength{\tabcolsep}{3.5pt}
  \caption{CER (\%) comparison on different subsets}
  \label{tab:cerdetail}
  \centering
  \begin{tabular}{cccccc}
   \toprule
    \textbf{Model}     &\textbf{Echoed}   &\textbf{$<$5 dB}  &\textbf{[5,15) dB}    &\textbf{$\geqslant$15 dB}  &\textbf{Total}\\
    \midrule
    Baseline  &16.49 &19.67 & 14.16 & 12.36 & 15.08 \\
    NNFB  &15.80 &18.61 & 13.82 & 12.45 & 14.67 \\
    DCUnet-MTL  & \textbf{14.68} & \textbf{16.11} & \textbf{12.54}   & \textbf{11.18} & \textbf{13.23} \\
    DCUnet-CAS  &15.11 &17.55 & 12.91   & 11.64 & 13.82 \\
    \bottomrule
  \end{tabular}\vspace{-6pt}
\end{table} 

The comparison of proposed and baseline systems on different subsets is shown in Table~\ref{tab:cerdetail}. DCUnet-MTL structure gives best performance on each subset. Compared with the NNFB method and baseline system, it shows 9.8\% and 12.2\% relative CER reduction totally. DCUnet-CAS brings sub-optimal results, but still shows advantages. Another conclusion drawn from the table is that the recognition rate of low SNR is still worse than that of high SNR. And even if there is a relatively small echo, it still has an impact on recognition accuracy.

\section{Conclusions}
\label{sec:typestyle}

With the belief that the advances of speech enhancement using complex spectrogram can benefit multi-channel ASR, in this study, we proposed two DCUnet-based methods for multi-channel  speech  recognition  and investigated some training tricks. Experiments on real-world 1000-hour XiaoMi smart speaker data with echos show that the proposed DCUnet approaches can bring superior speech recognition performance, while the DCUnet-MTL strategy outperforms the DCUnet-CAS strategy. Compared to the single-channel ASR model and the neural network fixed beamforming (NNFB) method, DCUnet-MTL can achieve 12.2\% and 9.8\% relative CER reduction, respectively. We notice that the complexity of our proposed models is high due to the use of many complex convolutional computations in DCUNet. In the future, we aim to reduce the complexity towards practical use and evaluate it on a larger data set.

\label{sec:ref}

\bibliographystyle{IEEEbib}
\bibliography{strings,refs}

\end{document}